\documentclass[aps,pra,twocolumn,superscriptaddress]{revtex4-1} 

\usepackage[caption=false]{subfig}
\usepackage{cancel} 
\usepackage{amsfonts}
\usepackage{amsthm,bm}
\usepackage{amsmath}
\usepackage{amssymb}
\usepackage{graphicx,xcolor}
\usepackage{float}
\usepackage{hyperref}
\usepackage{dcolumn}
\usepackage{soul}
\usepackage{ulem}
\usepackage{verbatim}

\begin{document}

\title{Confinement-induced unatomic trimer states }

\author{D. S. Rosa}

\affiliation{Instituto Tecnol\'{o}gico de Aeron\'{a}utica, DCTA,
  12228-900 S\~{a}o Jos\'{e} dos Campos, SP, Brazil}

\affiliation{Université Paris-Saclay, CNRS/IN2P3, IJCLab, 91405 Orsay, France}

\author{R. M. Francisco}

\affiliation{Instituto Tecnol\'{o}gico de Aeron\'{a}utica, DCTA,
  12228-900 S\~{a}o Jos\'{e} dos Campos, SP, Brazil}
  
  \author{T. Frederico}

\affiliation{Instituto Tecnol\'{o}gico de Aeron\'{a}utica, DCTA,
  12228-900 S\~{a}o Jos\'{e} dos Campos, SP, Brazil}

\author{G. Krein}

\affiliation{Instituto de F\'isica Te\'orica, Universidade Estadual Paulista,
Rua Dr. Bento Teobaldo Ferraz, 271-Bloco II, 01140-070 S\~ao Paulo, SP, Brazil}

\author{M. T. Yamashita}

\affiliation{Instituto de F\'isica Te\'orica, Universidade Estadual Paulista,
Rua Dr. Bento Teobaldo Ferraz, 271-Bloco II, 01140-070 S\~ao Paulo, SP, Brazil}

\begin{abstract}
The signature of an unatomic system is revealed by a continuous scale invariance that appears during a progressive dimensional squeezing of a resonantly interacting trimer. The unatomic regime is reached at the dimension $\overline D$, which for three identical atoms is found to 
be $\overline D=2.292${\textemdash}below this value, the trimer wave function at short distances displays a power-law behaviour. The fingerprint of this crossover is a sharp evolution of the contacts that characterizes the trimer momentum distribution tail.

\end{abstract}

\maketitle

\section{Introduction}

Symmetries play a fundamental role in various branches of physics, for example, in particle and nuclear physics, and statistical and condensed matter physics. In 2007, Georgi proposed~\cite{georgi} that beyond the Standard Model of elementary particles, there is a hidden symmetry sector of \textit{unparticles}. Unparticles can be created by local operators whose correlation functions display power-law behavior, with exponents determined by the scaling dimension of the fields in the operators.  Similarly to an elementary particle, which can be defined as an irreducible representation of the Poincar\'e group, an unparticle can be defined as an irreducible representation of the conformal symmetry group of a relativistic conformal field theory (CFT). The hypothesized unparticles can only be observed indirectly by measuring the recoil momentum distribution of Standard Model particles produced in association with them. 

Recently, Hammer and Son  considered a nonrelativistic analogue to the relativistic unparticle, which they called \textit{unnucleus}~\cite{HammerSon,commentHammerSon}. The unnucleus is defined as a field of a nonrelativistic CFT~\cite{nishidaCFT,mehenCFT}, characterized by a mass and a nonrelativistic conformal dimension. Differently from the relativistic unparticle, those authors argued that the unnucleus  exists in nature and can be realized in nuclear reactions involving neutron emission~\cite{unnucleus1,unnucleus2,unnucleus3}. Such a realization can occur when momentum $p$ is transferred with a length $\hbar/p$ lying between the range of the nuclear forces and the two-neutron scattering length.

It is possible to extend the definition made for particles and nuclei to include cold atoms. In this context, it is possible to realize \textit{unatomic} systems as the nonrelativistic counterpart of the relativistic CFT, which can be found as eigenstates of particular Hamiltonians e.g., the non-interacting ones. Although most Hamiltonians are not scale invariant, due to the presence of interactions, there  exists a way to have conformal symmetry in interacting theories by tuning the few-body system close to the unitarity limit or infinite scattering length. In general, it has been proposed to observe scaling symmetries in fermionic systems~\cite{fermions1,unnucleousatomic}, since systems composed by bosons can exhibit  discrete scale symmetry and the associated Efimov states~\cite{efimov0,efimov1}, which displays log-periodic behavior instead of a power-law one (for reviews, see Refs.~\cite{Braaten:2004rn,Naidon:2016dpf,Greene:2017cik,Hammer:2019poc}).

In this article, we argue that bound three-atom systems close to the unitarity limit can be driven into an unatomic regime by continuously deforming an atomic trap from three to two dimensions. The signature of the unatomic regime is characterized by a transition from  discrete to continuous scale symmetry. The transition occurs when the atomic trap drives the atomic cloud to a pancake shape~\cite{BEC3D,BEC2D,BEC1D} and the Feshbach resonance keeps the system with  a large scattering length and close to the unitarity limit~\cite{feshbach}.

In our work, we use a noninteger dimension ($D$) to mimic the deformation of the atomic trap.  For a three-atom system, the dimension $D$ can be related to an experimental setup with a harmonic well acting in the  deformation direction as~\cite{garridoconection3}:
\begin{equation}
b_{ho}^2/r_{2D}^2 = \frac{3(D-2)}{(3-D)(D-1)}\,,    
\end{equation}
where $b_{ho}$ is the oscillator length and $r_{2D}$ is the root-mean-square 
radius of the bound three-atom system in two dimensions ($D=2$). Recently, the relation was generalized to $N$-particles in arbitrary nonintegers dimensions below three dimensions ($D=3$)~\cite{garridoconection4}.

The Efimov effect for resonantly interacting three identical bosons occurs for values of $D$ in the range $2.3 < D < 3.8$~\cite{D1,mohapatra}, where the physics is driven by discrete scale symmetry. Furthermore, the three-body problem can be solved analytically in the unitary limit for arbitrary $D$ with the method introduced in Refs.~\cite{dsrSTM,betpeiPRA}. That method is also convenient for investigating the Efimov-unatomic transition below the critical dimension  $D_c = 2.3$, generalizing the treatment already used to study Efimov states. As we will show, the unatomic trimer is characterized by its wave function and momentum density.

This work is organized as follows.  For completeness and to make it self-contained, in Section~\ref{secII} we briefly review  the analytical derivation~\cite{dsrSTM,betpeiPRA} of the $D$-dimensional Faddeev components of a three-body bound-state wave function for three identical bosons. In section~\ref{secIII}, we identify and study the appropriated solutions for the scale parameter. Then, in Section~\ref{secIV}, we study the trimer shallow bound state wave function for the different regimes of discrete- and continuous-scale symmetry. In section~\ref{secV}, we compute the momentum distribution, discuss its high-momentum tail, from which we obtain the unatomic two- and three-body contacts. We present concluding remarks in Section~\ref{secVI}. The appendix gives the technical details of the large momentum sub-leading contributions to the single particle momentum distribution in the unatomic regime.

\section{Trimer at the unitarity limit}
\label{secII}

We describe the trimer as an eigenstate of the free Hamiltonian with pairwise contact interaction and impose the Bethe-Peierls (BP) boundary condition~\cite{bethe}. The trimer wave function $\Psi(\textbf{x}_{1},\textbf{x}_{2},\textbf{x}_{3})$  is totally symmetric in the coordinates $\textbf{x}_{1}, \textbf{x}_{2}, \textbf{x}_{3}$ and can be written as the sum of Faddeev components $\psi(\mbox{\boldmath$r$}_i,\mbox{\boldmath$\rho$}_i)$:
\begin{equation}\label{eq:fullwavefunct}
\Psi(\textbf{x}_{1},\textbf{x}_{2},\textbf{x}_{3}) = \sum_{j=1}^3
\psi(\mbox{\boldmath$r$}_j,\mbox{\boldmath$\rho$}_j) \, ,
\end{equation}
where $\mbox{\boldmath$r$}_j$ and $\mbox{\boldmath$\rho$}_j$ are relative Jacobi coordinates given in terms of the atoms coordinates $\textbf{x}_i$ as
\begin{equation}
 \mbox{\boldmath$r$}_{i} = \textbf{x}_{j} - \textbf{x}_{k}\quad\text{and}\quad
 \mbox{\boldmath$\rho$}_{i} = \textbf{x}_i - \frac{1}{2}(\textbf{x}_j + \textbf{x}_k) \, ,
\end{equation}
with ($i, j, k$) taken cyclically among ($1,2,3$). Each Faddeev components $\psi(\mbox{\boldmath$r$}_i,\mbox{\boldmath$\rho$}_i)$ is an eigenstate of the free Hamiltonian. For convenience, we introduce the scaled 
coordinates $\overline{\textbf{r}}_{i} = \sqrt{\eta}\, \textbf{r}_{i}$ and $\overline{\bm{\rho}}_{i} = \sqrt{\mu} \bm{\rho}_{i}$ with reduced masses given by
\begin{equation}
  \eta =m/2\quad\text{and} \quad \mu = 2m/3\,.
  \label{eq:reducecmass}
\end{equation}
The Jacobi relative momenta written in terms of the momenta canonically conjugated to the scaled coordinates are given by
\begin{equation}\label{eq:rescmomenta}
\textbf{p}={\sqrt{\eta}}\,\overline{\textbf{p}}\quad\text{and}\quad \textbf{q}=\sqrt{\mu}\,\overline{\textbf{q}}\,,
\end{equation}
where $\textbf{p}$ is the relative momentum of one pair and $\textbf{q}$ is the relative momentum of the third particle to the center-of-mass of the pair. Note that we have three sets of scaled coordinates that are related to each other by an orthogonal transformation:
\begin{eqnarray}
\overline{\bm{r}}_{j}& =& - \overline{\bm{r}}_{k}\cos\theta + \overline{\bm{\rho}}_{k}\sin \theta \nonumber \\
\overline{\bm{\rho}}_{j}& =& - \overline{\bm{r}}_{k}\sin\theta - 
\overline{\bm{\rho}}_{k}\cos \theta,
\end{eqnarray}
where $\tan \theta =\sqrt{3}$. Using hyperspherical coordinates $(R,\alpha)$ such that $\overline r = R \sin \alpha$ and $\overline{\rho} = R \cos \alpha $, with $R^2 = \overline{r}^2+\overline{\rho}^2$ and $\alpha = \arctan{(\overline{r}/\overline{\rho})}$, the Faddeev component of the wave function for a given trimer bound-state energy, $E_3 = - \hbar^2\kappa_0^2/m$,
is given by~\cite{betpeiPRA}:
\begin{equation}
\psi(R,\alpha) =   K_{ s_n}\big(\sqrt{2} \kappa_0 R 
\big) \, \frac{\mathcal{T}(\alpha,s_n)}
{ R^{D-1}},
\label{wavefunction}
\end{equation}
where $K_{ s_n}$ is the modified Bessel function of the second kind,  $s_n$ is the  scale parameter  and
\begin{multline}
\mathcal{T}(\alpha,s_n) \equiv \frac{\sqrt{\sin2 \alpha}}{\left(\cos \alpha\ \sin \alpha\right)^{\frac D2-\frac12}}
\times\Bigg\{ P_{\frac{s_n}{2}-\frac12}^{\frac D2-1}\left(\cos2 \alpha \right) \\ - \frac{2}{\pi}\tan\left[\frac\pi2(s_n-1)\right] Q_{\frac{s_n}{2}-\frac12}^{\frac D2-1}\left(\cos2 
\alpha\right)
\Bigg\}\, ,
\label{angular}
\end{multline}
where  $P_{n}^{m}(x)$ and $Q_{n}^{m}(x)$ are the associated Legendre functions.  

The BP boundary condition in the situation that three boson pairs interact resonantly leads to a characteristic equation determining the scale parameter $s_n$, given by
\begin{multline}
\frac{1}{2}
\left[ \left(\cot\alpha\right)^{\frac D2-\frac12} 
\left( \sin2\alpha \frac{\partial}{\partial \alpha} 
+ D-3\right) G (\alpha) \right]_{\alpha\rightarrow 0}\\
+ 2(D -2) \frac{ G(\theta)}
{\left(\sin\theta \cos\theta\right)^{\frac D2-\frac12}}  = 0\,,
\label{scale}
\end{multline} 
where 
\begin{multline} 
G(\alpha) = (\sin 2{\alpha})^\frac12 
\left\{P_{\frac{s_n}{2}-\frac12}^{\frac D2-1}(\cos 2{\alpha}) \right.  \\ -\left. \frac{2}{\pi} \tan\left[ {\frac{\pi(s_n - 1)}{2}}\right] Q_{\frac{s_n}{2}-\frac12}^{\frac D2-1}(\cos{2{\alpha}})\right\}\,
\end{multline}
satisfies the hyperspherical differential equation that arises from the separation of the hyperspherical variables~\cite{betpeiPRA}
\begin{equation}
\left[-\frac{\partial^{2}}{\partial \alpha^{2}} -s_n^{2} + \frac{(D-1)(D-3)}{\sin^{2}2{\alpha}}\right]G({\alpha}) = 0.
\end{equation} 

\section{D-DIMENSIONAL SCALE PARAMETERS}
\label{secIII}

In the range $D_c < D \leq 3$, where $D_c$ denote the critical dimension, the characteristic equation~\eqref{scale} has an infinite number of real solutions and only one imaginary ($s_n =\pm i s_0$), being the latter one responsible for the attractive Efimov effective potential. For $D=3$, this solution reduces to the Efimov discrete scale parameter  $s_0=1.00624...$. At the critical dimension, the only solution is $s_n = 0$. Below $D_c$, that is given by $D_c = 2.3$ for three identical bosons~\cite{D1}, we also find an infinite number of real solutions. However, as we will see below, not all of these solutions allow the normalization of the wave function. To constrain the solutions of the characteristic equation for $2 < D < D_c$, we demand consistence with the solution of the Skorniakov and Ter-Martirosyan (STM) integral equation in noninteger dimensions~\cite{dsrSTM} for the dimer bound at the scattering threshold. We require that the spectator function derived from the Faddeev component of the trimer wave function in Eq.~\eqref{wavefunction} solves the trimer STM equation in $D$-dimensions. The spectator function in momentum space, which we denote as $\chi(q)$, is the solution of the STM equation in $D$ dimensions
\begin{align}
\chi(q)&=\frac{2}{\pi^{D/2}} \Big(\frac43\Big)^{\frac D2-1}
{\Gamma\Big(\frac D2\Big)} \nonumber \\
&\times q^{2-D}\int d^{D}k \frac{\chi(k)}{ \kappa_0^{2}+q^{2}+k^{2}+\textbf{q} \cdot \textbf{k} }\,,
\label{stmkernel}
\end{align}
where  $\textbf{k}$ and $\textbf{q}$ are the Jacobi relative  momenta of one of the spectator particles with respect to the center-of-mass of the other two.  $\chi(q)$ was derived from the Faddeev component of the trimer wave function in the configuration space in Ref.~\cite{Nb_AAB_Ddim_Efimov} and is briefly reviewed in the following. 

The starting point is the Faddeev wave function in Eq.~\eqref{wavefunction}, which, for small distances, is given by:
\begin{equation}
\psi(\overline{\textbf{r}},\overline{\bm \rho})|_ {\overline{r}\to 0}\propto
\frac{K_{s_n}\left(\sqrt{2}\,\kappa_0\, \overline{\rho}\right)}{\overline{\rho}\;\overline{r}^{D-2}}\,. 
\label{wfrlimit}
 \end{equation}
The wave function $\psi(\overline{\textbf{r}},\overline{\bm \rho})$ obeys 
the Schr\"odinger’s equation with contact interaction~\cite{Nb_AAB_Ddim_Efimov}
\begin{equation}
\left[\nabla^{2}_{\overline{\textbf{r}}} +
\nabla^{2}_{\overline{\bm{\rho}}}+2 \kappa_0^2\right] \psi(\overline{\textbf{r}},\overline{\bm{\rho}})= \delta(\overline{\textbf{r}})B(\overline{\bm{\rho}})\,,
\label{freeschroe}
\end{equation}
where $B(\overline{\rho})$ is the spectator function in configuration space. 
Substituting Eq.~\eqref{wfrlimit} in~\eqref{freeschroe}, we find
\begin{eqnarray}
B(\overline{\rho})&\propto&
\frac{K_{s_n}\left(\sqrt{2}\kappa_0 \overline{\rho}\right)}{\overline{\rho}}\,.
 \label{specfuncrho}
\end{eqnarray}
The spectator function $\chi(q)$ is the $D$-dimensional Fourier transform 
\begin{equation}
\chi(q)=\int d^{D}\overline{\rho}\, \exp\left(-i \,{\frac{\textbf{q}}{\sqrt{\mu}}}\cdot\overline{\bm{\rho}}\right)B(\overline{\rho})  \,,
\label{specposition}
\end{equation}
where the factor $1/\sqrt{\mu}$ comes from the scaling of the momentum $\overline{\bm{q}}$ with respect to the relative Jacobi momentum (see Eq.~\eqref{eq:rescmomenta}). 
The explicit solution for $\chi(q)$ is given by
 \begin{eqnarray}
 \chi(q)\propto
H_2  \tilde{F}_1 \left(\mathcal{F}_{(D,s_n)}^{-} ,\mathcal{F}_{(D,s_n)}^{+},\frac{D}{2},-\frac{{q}^{ 2}}{2\mu\kappa_0^2}
 \right),
 \label{regulatespect}
\end{eqnarray} 
where $H_2 \tilde{F}_1(a,b,c,z)$ is the regularized hyper-geometrical function and
$\mathcal{F}_{(D, s_n)}^{\pm} \equiv (D-1\pm s_n)/{2}$. 
Equation~\eqref{regulatespect}
is the exact solution of the $D-$dimensional STM equation, Eq.~\eqref{stmkernel}. For $2<D<D_c$, where $s_n$ is real, only a subset of the solutions of the characteristic equation~\eqref{scale} leads to a finite integral in the STM equation. This can be shown by inspecting the  high momentum tail of the spectator function
\begin{equation}
\chi(q)\underset{q\gg \kappa_0}{\propto}
q^{1-D}\Big[ \mathcal{G}_{+ s_n} q^{+s_n}
+   \mathcal{G}_{- s_n} 
q^{-s_n}
\Big],
\label{asympespec}
\end{equation}
where
\begin{eqnarray}
 \mathcal{G}_{\pm s_n}  
\equiv\frac{\Gamma(\pm s_n)}{\Gamma\left[(D-1\pm s_n)/2\right]\Gamma\left[(1\pm s_n)/2\right]}\, .
 \label{eq:G}
\end{eqnarray}
Power counting shows that the integral in Eq.~(\ref{stmkernel}) converges for  $D-2+1-D\pm s_n<0$, so that, below $D_c$, the scale parameter have to be constricted in the interval $-1<s_n<1$.   Since Eq.~\eqref{asympespec} is symmetric under $s_n \rightarrow -s_n$, from now on we use $s_1$ to denote the real values of $s_n$ in the interval $-1 < s_1 < 0 $ and use the positive values $s_0 > 0$ in the Efimov regime, where $s_n = is_0$.

\begin{center}
\begin{figure}[h]
\includegraphics[width=8.5cm]{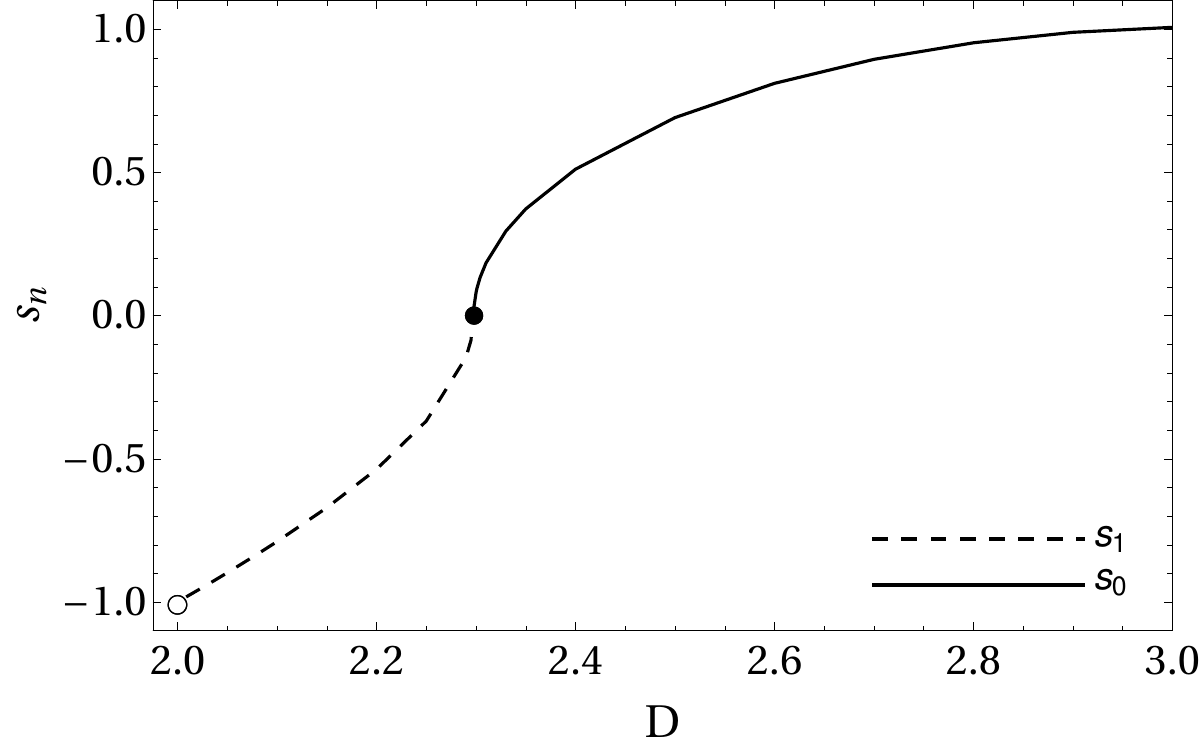}
\caption{Trimer scale parameter $s_n$ dependence on the dimension~$D$. The dashed line represents the real physical solutions ($s_1$) in the unatomic regime, while the solid line represents the imaginary roots ($s_0$) in the Efimov regime. The black point indicates the critical dimension $D_c=2.3$ and the empty circle corresponds to the limit $D\to2_+$.}
\label{fig1}
\end{figure}
\end{center}

Figure~\ref{fig1} displays the $D$ dependence of the solutions $s_n$ of the characteristic Eq.~\eqref{scale}. The black point in the figure denotes the critical dimension, $D_c=2.3$, where $s_n = 0$. For $D_c < D \leq 3$, the values of $s_0$ in the imaginary solutions ($s_n = i s_0$) are shown, whereas for $2 < D \leq D_c$ the real values $s_n = s_1$ are displayed.  

Finally, note that the asymptotic power law form given in Eq.~\eqref{asympespec} is valid for dimensions strictly larger than two and smaller than {four~\cite{Nb_AAB_Ddim_Efimov}, since for $D=2$,
$\chi(q)\to q^ {-2}\log q$ for $q\to \infty$~\cite{Bellotti:2012dv}; the limit $D > 2$ is indicated by the empty black circle in Fig.~\ref{fig1}}. {For a contact interaction in $D=2$, the STM equation 
demands a finite dimer energy, and} the trimer energy and its structural properties  scale with that  energy. Therefore, our solution of the STM equation at the unitary limit cannot be used to represent the trimer in $D=2$.

\section{Unatomic Trimer wave function}
\label{secIV}

\begin{figure}[h]
\includegraphics[width=8.5cm]{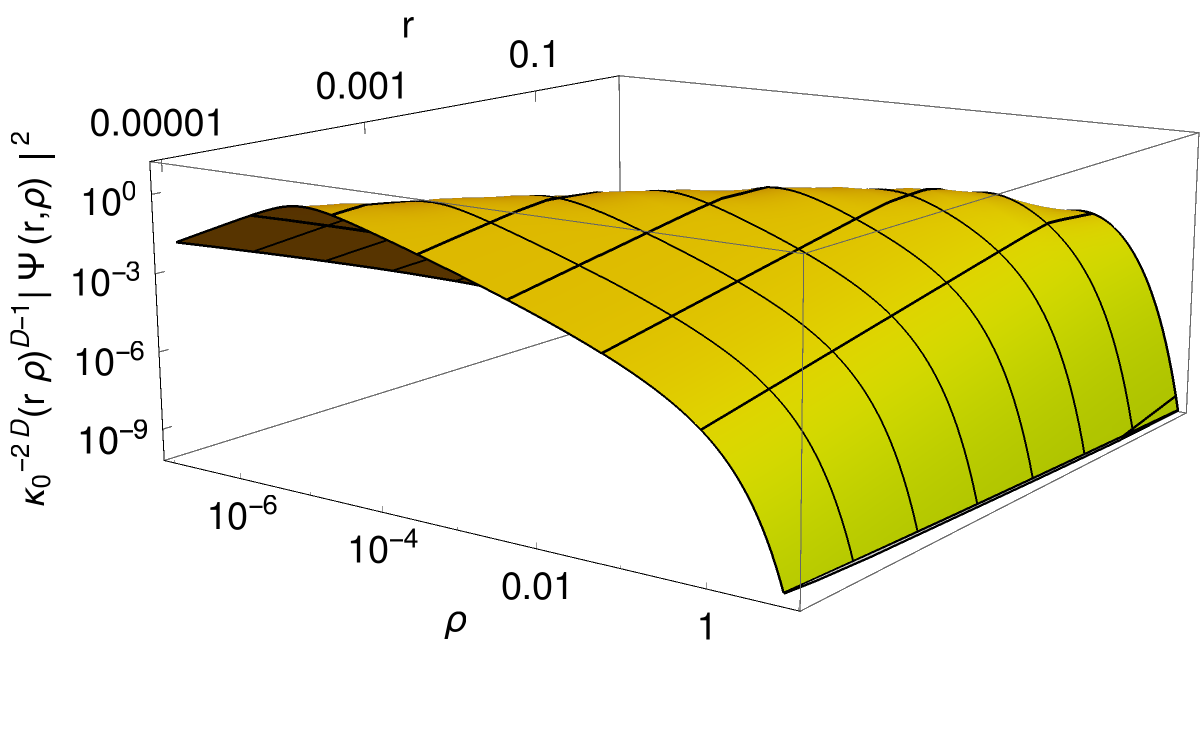}
\includegraphics[width=8.5cm]{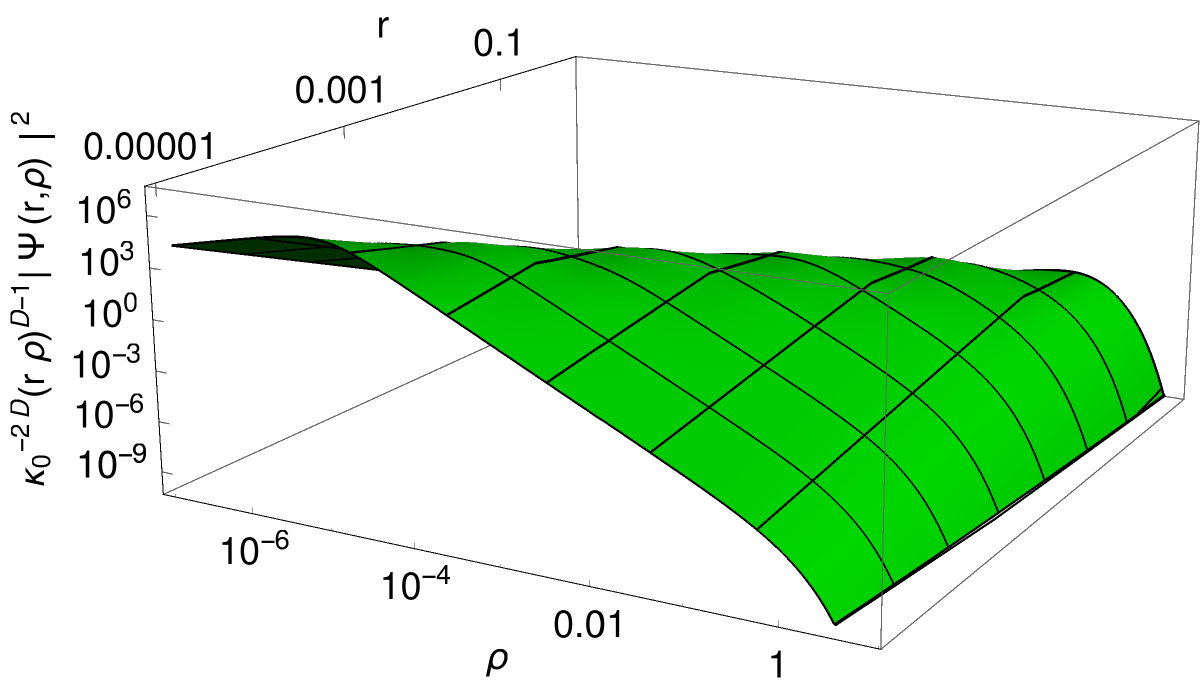}
\caption{ Dimensionless radial distribution from the trimer wave function $\Psi$, Eq.~\eqref{eq:fullwavefunct},  as a function of dimensionless quantities $r=\kappa_0 r_3$ (pair  relative distance) 
and $\rho=\kappa_0 \rho_3$ (third atom relative distance to the pair system).  The angle between $\bm{r}$ and $\bm\rho$ is fixed to $\pi/3$. Upper panel: results for the critical dimension $D_c =2.3$. Lower panel: results for $D=2.1$.} 
\label{fig2}
\end{figure}

The following dimensionless radial distribution will be used to illustrate some of the unatomic properties 
\begin{equation}\label{eq:radialdist}
\kappa_0^{-2D}(r\,\rho)^{D-1}|\Psi({\bm r},{\bm\rho})|^2\,.
\end{equation}
We start by showing, in the upper panel of Fig.~\ref{fig2}, results for the total trimer wave function for the system with $D = D_c$, for which $s_n=0$ and  $b_{ho}/r_{2D}=0.9945$. The angle between ${\bm r}$ and ${\bm \rho}$ is fixed to $\pi/3$. In the lower panel of Fig.~\ref{fig2}, we  present the wave function for $D=2.1$, for which $s_n\equiv s_1=-0.788$ and $b_{ho}/r_{2D}=0.55$.  One notes that the typical Efimov nodes are no longer present for values less than or equal to the critical dimension $D_c =2.3$. In addition, the unatomic curve, lower panel, is up to six orders of magnitude larger than the curve in the upper panel, a feature that can be understood by the fact that in this case the system is closer to the three-body threshold ($D=2$), where, in the unitary limit, the three-body binding energy is proportional to the two-body one.

In what follows, we derive the functional forms of the trimer wave functions for the different limits associated with the scale parameter. This will reveal the signature of unatomic states. 

The trimer shallow bound state wave function for the different regimes of discrete and continuous scale symmetry is defined by the values of $s_n$. Each Faddeev component, Eq.~\eqref{wavefunction}, in the limit of $ \kappa_0 R\to 0$, is given by
\begin{equation} \psi(R,\alpha)|_{\kappa_0 R\to 0}\propto \frac{\mathcal{T}(\alpha,s_n)}{R^{D-1}}\Lambda(R,s_n)  \, , \label{wavefunctionlimit0} \end{equation}
where 
\begin{equation}\label{eq:Lambda}
    \Lambda(R,s_n) =\kappa_0^{s_n} \frac{\Gamma(-s_n)}{2^{s_n/2}}R^{s_n} 
+
\frac{\kappa_0^{-s_n}\Gamma(s_n)}{2^{-s_n/2}}R^{-s_n} \, .
\end{equation}
For $2<D<D_c$,  $s_n\equiv s_1$ with $-1<s_1<0$, so that only one of the terms in Eq.~\eqref{eq:Lambda} dominates at short distances, leading to

\begin{eqnarray}
\psi(R,\alpha)|_{\kappa_0 R\to 0}\propto \frac{\mathcal{T}(\alpha,s_1)}{R^{D-s_1-1}}\,,
\label{wavefunctionlimit1}
\end{eqnarray}
 which behaves as a homogeneous function with a non-trivial exponent determined by the noninteger dimension. This means that it is invariant under a continuous scale transformation, namely 
\begin{equation}
\psi(\lambda\overline{\bm r},\lambda \overline{\bm \rho}) \propto \psi(\overline{\bm r},\overline{\bm \rho})\,.
\end{equation}
This property is trivially valid for the trimer wave function, Eq.~\eqref{eq:fullwavefunct}, and is the signature of system's unatomic nature. For comparison,  the Faddeev component of the trimer wave function in the Efimov regime, where $s_n \equiv i s_0$, derived from Eq.~\eqref{wavefunctionlimit0}, is given by
\begin{equation}
\psi(R,\alpha)|_{\kappa_0 R\to 0}\propto
 \cos\left[ s_0 \ln \left( \frac{\kappa_0R}{\sqrt{2}}\right) - \Phi \right] \frac{\mathcal{T}(\alpha,i s_0)}{R^{D-1}}\,,
\label{wavefunctionlimitefimov}   
\end{equation}
where the phase 
\begin{equation}
\Phi  \equiv\pi+\arctan\{\operatorname{Im}[\Gamma(is_0)]/\operatorname{Re}[\Gamma(is_0)]\}\, 
\end{equation}
also depends on the noninteger dimension. In constrast to the unatomic case, Eq.~\eqref{wavefunctionlimitefimov}  is invariant under discrete scale  transformation 
\begin{equation}
\psi\left(e^{n\pi/s_0}\overline{\bm r},e^{n\pi/s_0}\overline{\bm \rho}\right) \propto \psi(\overline{\bm r},\overline{\bm \rho})\,,
\end{equation}
with $n$ integer. 

At the critical dimension $D_{c}$, we have $s_n \equiv 0$, so that, in the trimer shallow bound-state limit, $\Lambda(R,s_n)$ reduces to:
\begin{equation}   \Lambda(R,s_n)|_{s_n\to0} = 2\ln\left(\frac{2e^{-\gamma}}{\kappa_0 R}\right)\,.
\end{equation}
Therefore, the Faddeev component of the trimer wave function~\eqref{wavefunctionlimit1} in that limit becomes
\begin{eqnarray}
\psi(R,\alpha)|_{\kappa_0 R\to 0}&\propto  &
\frac{\ln\left(2e^{-\gamma}/\kappa_0 R\right)}{R^{D_c-1}}\,\nonumber \\
 &\times& \frac{Q_{-\frac{1}{2}}^{\frac{D_c}{2}-1}(\cos2\alpha)\,\sqrt{\sin2\alpha}}{(\cos2\alpha \sin2\alpha)^\frac{D_c-1}{2}}.
\label{wavefunctionlimits0}
\end{eqnarray}

\begin{figure} \includegraphics[width=8.5cm]{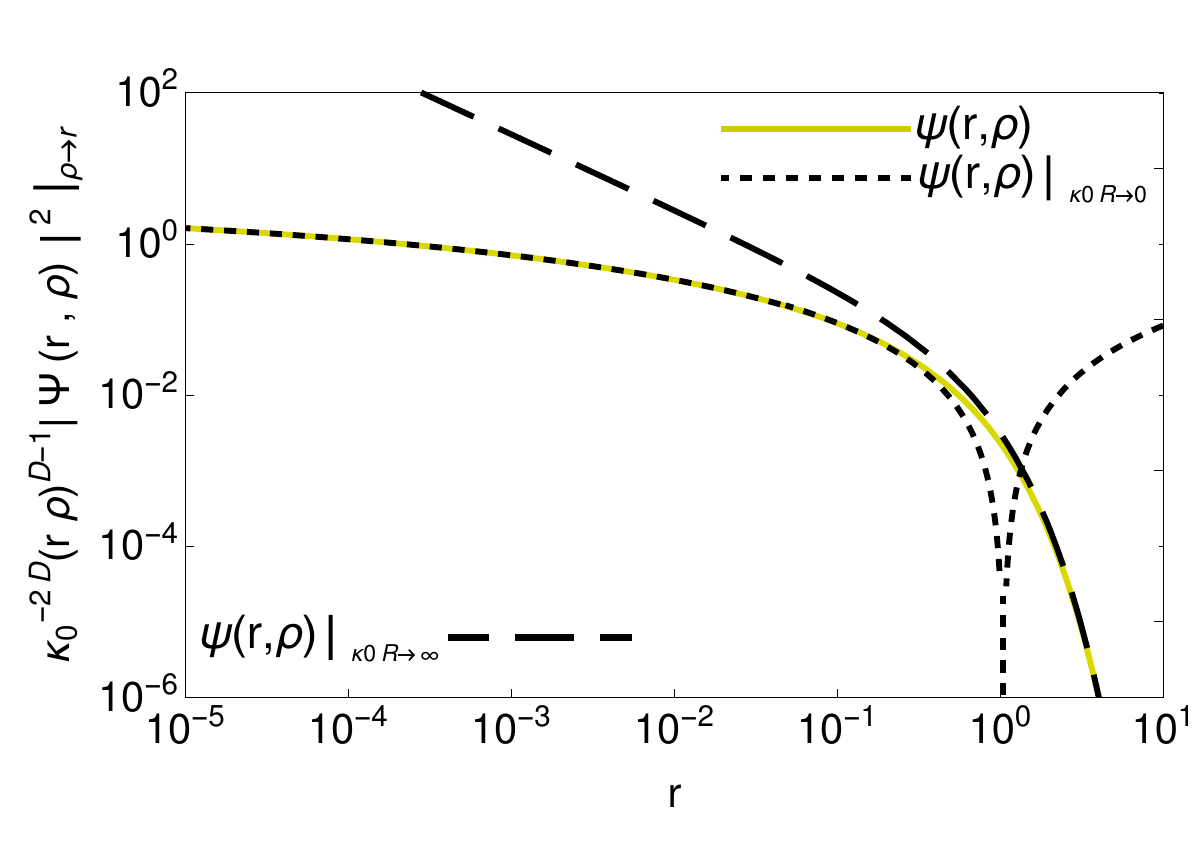}
\includegraphics[width=8.5cm]{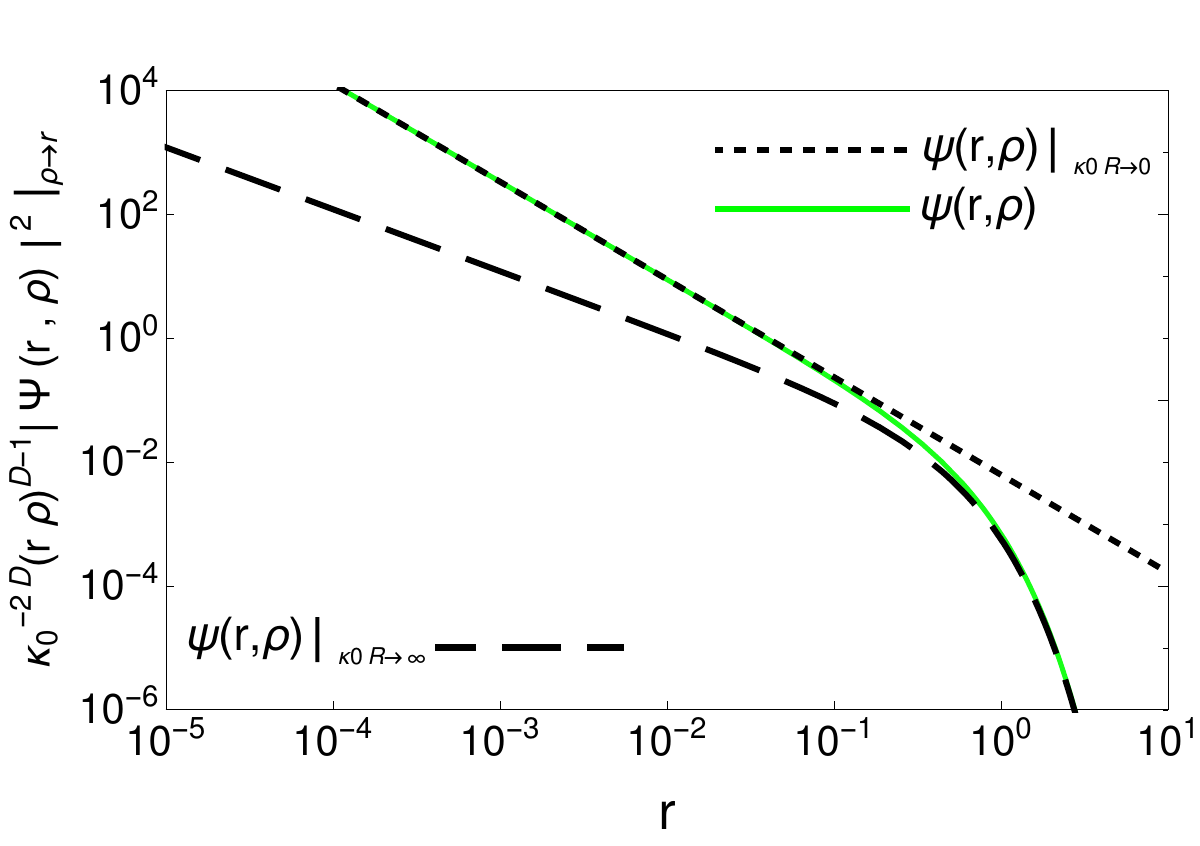} \caption{Dimensionless probability distribution as a function of dimensionless quantity $r\equiv \kappa_0r_3$ for $r_3=\rho_3$. We contrast the asymptotic formulas for small and large distances. Upper panel: trimer wave function for $D=D_c$, Eq.~\eqref{eq:radialdist} (solid line), small distances trimer wave function, Eq.~\eqref{wavefunctionlimits0} (dotted line), and large distances trimer wave function, Eq.~\eqref{wavefunctiontail} (dashed line). Lower panel: trimer wave function for $D=2.1$ (with $b_{ho}/r_{2D}=0.55$), Eq.~\eqref{eq:radialdist} (solid line), small distances trimer wave function, Eq.~\eqref{wavefunctionlimit1} (dotted line), and
large distances trimer wave function, Eq.~\eqref{wavefunctiontail} (dashed line). The angle between ${\bm r}$ and ${\bm\rho}$ was fixed to $\pi/3$. } \label{fig3} 
\end{figure}

For finite $\kappa_0$ and $r=\rho$, figure~\ref{fig3} shows in solid lines the trimer dimensionless distribution given by Eq.~\eqref{eq:radialdist}, with the angle between ${\bm r}$ and ${\bm\rho}$ equal to $\pi/3$. Also, dotted and dashed lines display the numerical results for the asymptotic formulas for small and large distances, respectively. For small distances, the asymptotic formula is given by Eq.~\ref{wavefunctionlimit0}, while for large distances it is given by
\begin{equation}
\psi(R,\alpha)|_{\kappa_0R\to\infty}\to
 \frac{ \exp\left(-\sqrt{2} \kappa_0 R
\right) }
{ R^D}\mathcal{T}(\alpha,s_n)\,,
\label{wavefunctiontail}
\end{equation}
which is suitable for any  noninteger dimension in the interval $2<D\leq3$. 

The upper panel of Fig.~\ref{fig3} shows the distribution computed with the exact trimer wave function for $D=D_c$, which is built with the Faddeev components from Eq.~\eqref{wavefunction}. In this frame this function is compared to the small distances asymptotic formula given by Eq.~\eqref{wavefunctionlimits0}, which approximates well the exact result for $\kappa_0r$ below 0.1. The asymptotic formula for large distances, derived from Eq.~\eqref{wavefunctiontail}, works well for $\kappa_0r$ larger than 1.

In the lower panel of Fig.~\ref{fig3}, we plotted the trimer exact distribution for $D=2.1$ $(b_{ho}/r_{2D}=0.55)$ (solid line). It is compared to the small distances distribution built with the power-law Faddeev component of the wave function, Eq.~\eqref{wavefunctionlimit1}, which approximates well the exact distribution for $\kappa_0r$ below 0.1. The large distance distribution, given by Eq.~\eqref{wavefunctiontail}, works well for $\kappa_0r$ above unity. It is observed that the unatomic character of the trimer is attained for  $R\lesssim\kappa_0^{-1}$, i.e., smaller than the typical size of a bound trimer, given by $1/\kappa_0$.

\section{Momentum distribution}
\label{secV}

In this section, we will follow closely the procedure of reference~\cite{Nb_AAB_Ddim_Efimov}, where the momentum distribution for mass imbalanced systems in the Efimov regime was calculated. Both Efimov and unatomic regimes are driven by asymptotic behaviors of the total wave function,  depending on the effective dimension, momentum and energy. 

The consequences of the power-law behavior in the Faddeev component of the trimer state can be studied investigating the high momentum tail of the single particle momentum distributions, such that it is possible to define an effective dimension $\overline D < D_c$ in which the unatomic behavior is highlighted and starts to dominate the dynamics of the system. 

The single particle momentum distribution is computed from the momentum space representation of the trimer wave function. In this procedure, the Faddeev component of the wave function in momentum space representation is obtained from the spectator function, $\chi(\textbf{q})$, which was already computed in equation~\eqref{regulatespect}. The total trimer  wave function in momentum space, namely the Fourier transform of Eq.~\eqref{eq:fullwavefunct}, is the sum over the Faddeev components
\begin{equation} \label{eq:wfp}
\hspace{-.2cm}
\langle \textbf{q}\,, \textbf{p} | \Psi \rangle = \frac{\chi(|\textbf{q}|)  + \chi\left(\bigl\vert\textbf{p} - {\textbf{q}}/{2}\bigr\vert\right)
+ \chi\left(\bigl\vert\textbf{p} + \textbf{q}/2\bigr\vert\right) }{\kappa_0^2+p^2 + \frac{3}{4}q^2 },
\end{equation}
where we have the standard Jacobi relative momentum, not the scaled ones, given by Eq.~\eqref{eq:rescmomenta}. 

In the range of dimensions $2<D<D_c$, where $s_n=s_1$ with $-1<s_1<0$, the large momentum behaviour ($q\gg\kappa_0$) of the spectator function, Eq.~\eqref{asympespec}, is given by the homogeneous function
 \begin{equation}
\chi(q)\underset{q\gg \kappa_0}{\propto}
 q^{1-D-s_1}
 \,.
\label{asympespectunatomic}
\end{equation} 
Thus, the  unatomic trimer wave function at large momenta is invariant under continuous scale transformations as 

\begin{equation}\label{eq:unatomictail}
 \langle \lambda \textbf{q}\,, \lambda\textbf{p} | \Psi \rangle =\frac{1}{\lambda^{D-1+s_1}}\langle\textbf{q}\,, \textbf{p} | \Psi \rangle \,,
\end{equation}
where the scaling coefficient, $-(D-1+s_1)$, is always negative in the range $2<D<D_c$ as $1+s_1>0$.

From now on, we focus on the normalized single particle momentum distribution, which is computed from
\begin{equation}
\hspace{-.1cm}
 n(q) = \int d^{D}p  \mid\langle \textbf{q}, \textbf{p} \mid \Psi \rangle \mid^{2}~ \text{and}~ \int d^Dq\, n(q) =1\,,
\label{densityB}
 \end{equation} 
where the trimer bound state wave function in momentum space is given by Eq.~\eqref{eq:wfp}.

The momentum density can be separated
into nine terms, which by considering the symmetry of the wave function is reduced to four. The calculation of the momentum
density is then simplified to the determination of the following terms:
 \begin{equation}
 \label{4sum}
 n(q) = n_1(q) + n_2(q) + n_3(q) + n_4(q),
 \end{equation}
which are given by
 \begin{eqnarray}
&&n_{1}(q)= \lvert \chi(q) \rvert^{2} \int d^{D}p \frac{1}{\left(\kappa_0^2 + 
 p^{2}+\frac{3}{4}q^{2} \right)^{2}}, \label{n1} \\
&&n_{2}(q) = 2 \int d^{D}p \frac{\lvert \chi(\lvert \textbf{p} -\textbf{q}/2 \rvert) \rvert^{2}}
 { \left( \kappa_0^2 + p^{2}+\frac{3}{4} q^{2} \right)^{2} }, \label{n2} \\
&&n_{3}(q) = 2  \chi^{*}(q) \int d^{D}p \frac{\chi(\lvert \textbf{p} - \textbf{q}/2  \rvert )  }
 { \left( \kappa_0^2 + p^{2} +  \frac{3}{4}q^{2}  \right)^{2} }+ {\rm c.c.}, \label{n3}\\
&&n_{4}(q) = \int d^{D}p \frac{\chi^{*}(\lvert \textbf{p}-\textbf{q}/2 \rvert)
 \chi(\lvert \textbf{p} + \textbf{q}/2  \rvert )  }
 {\left( \kappa_0^2 + p^{2} +\frac{3}{4} q^{2}   \right)^{2}} + {\rm c.c.}.
\ \ \ \ \
 \label{n4}
 \end{eqnarray}

\begin{center}
\begin{figure}
\includegraphics[width=8.5cm]{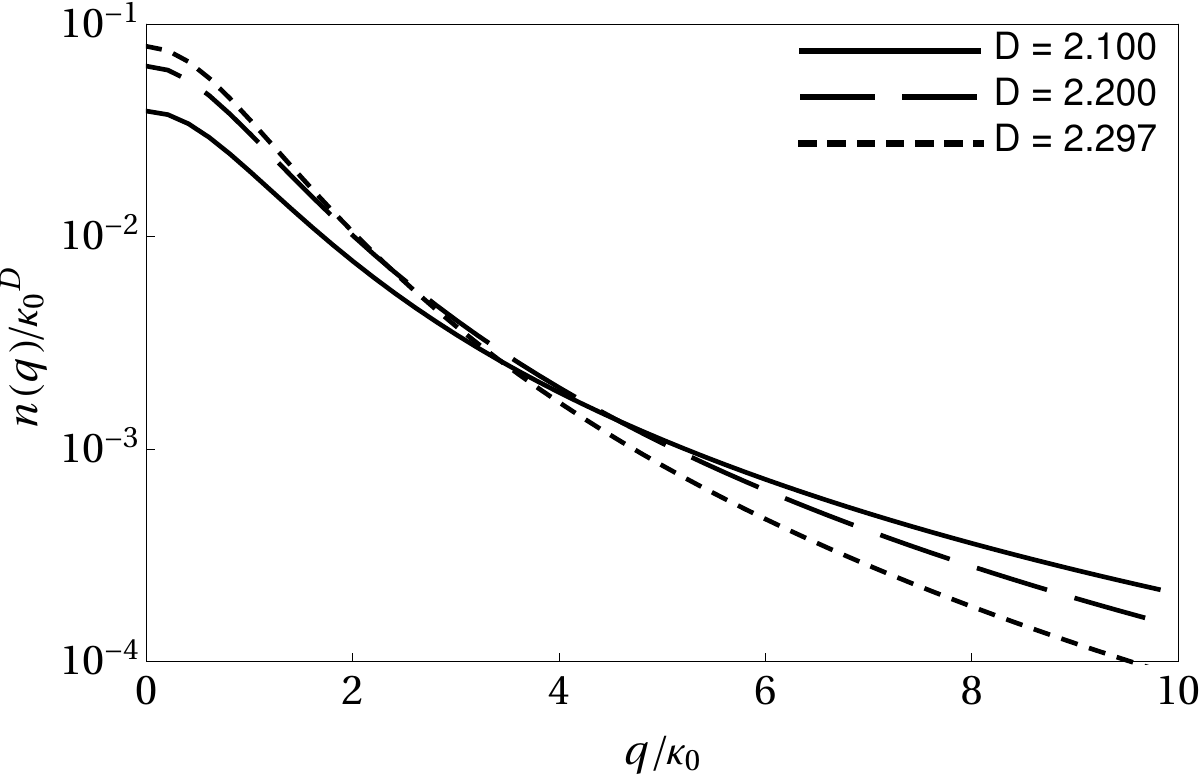}
\caption{Single-particle momentum distribution for three identical atoms embedded in noninteger dimensions. Results obtained with the regular spectator function Eq.~\ref{regulatespect}. }
\label{fig4}
\end{figure}
\end{center}

The  momentum density, $n(q)$, is shown in Fig.~\ref{fig4}, for the noninteger dimensions $2.297$, $2.2$ and $2.1$, where we see the enhancement of the high momentum tail by decreasing the noninteger dimension. A similar effect was also found for the Efimov regime in~\cite{Nb_AAB_Ddim_Efimov}. A naive physical interpretation of this increase can be made considering that the suppression of one spatial dimension should emphasize the large momentum tail by the uncertainty principle.

The investigation of the high-momentum region of the momentum distribution returns some important universal quantities, namely the contact parameters, which are used to parameterize thermodynamic relations between macroscopic observables~\cite{Tan1,Tan2,Tan3}. Following the steps detailed in App.~\ref{app:a}, the leading contribution below the critical dimension can be derived as
\begin{eqnarray} \label{eq:nasympdbardc}
n(q) \underset{q  \gg \kappa_0}{=} \frac{{C}_{2}}{q^{4}}+\frac{{C}_{3}}{q^{D+2+2s_1}} + \cdots, ~ \text{for}~ \overline D<D<D_c\,,
\end{eqnarray}
and
\begin{eqnarray}
n(q) \underset{q  \gg \kappa_0}{=} \frac{{C}_{3}}{q^{D+2+2s_1}} + \cdots, ~ \text{for}~  2< D<\overline D,
\label{c2c3}
\end{eqnarray} 
where $C_2$ and  ${C}_3$ are the two and three-body contact parameters, respectively. Here, $\overline{D}$ represents the dimension in which the asymptotic behavior of the momentum distribution becomes the one given in equation~\ref{c2c3}, remaining in this way until  $D=2$, where the system reaches the three-body threshold. The two and three-body contacts have dimension of $(\text{length})^{-4}$ and
$(\text{length})^{-2-2s_1}$. They scale as
${C}_2\propto \kappa_0^{4}$ and
${C}_3\propto \kappa_0^{2+2s_1}$. The source of the two-body contact contribution to the density is only due to $n_2(q)$, as discussed in App.~\ref{app:a} and in references~\cite{castindensity,braaten,yamashita2013,braaten2014,Nb_AAB_Ddim_Efimov}.

\begin{figure}
\includegraphics[width=8.5cm]{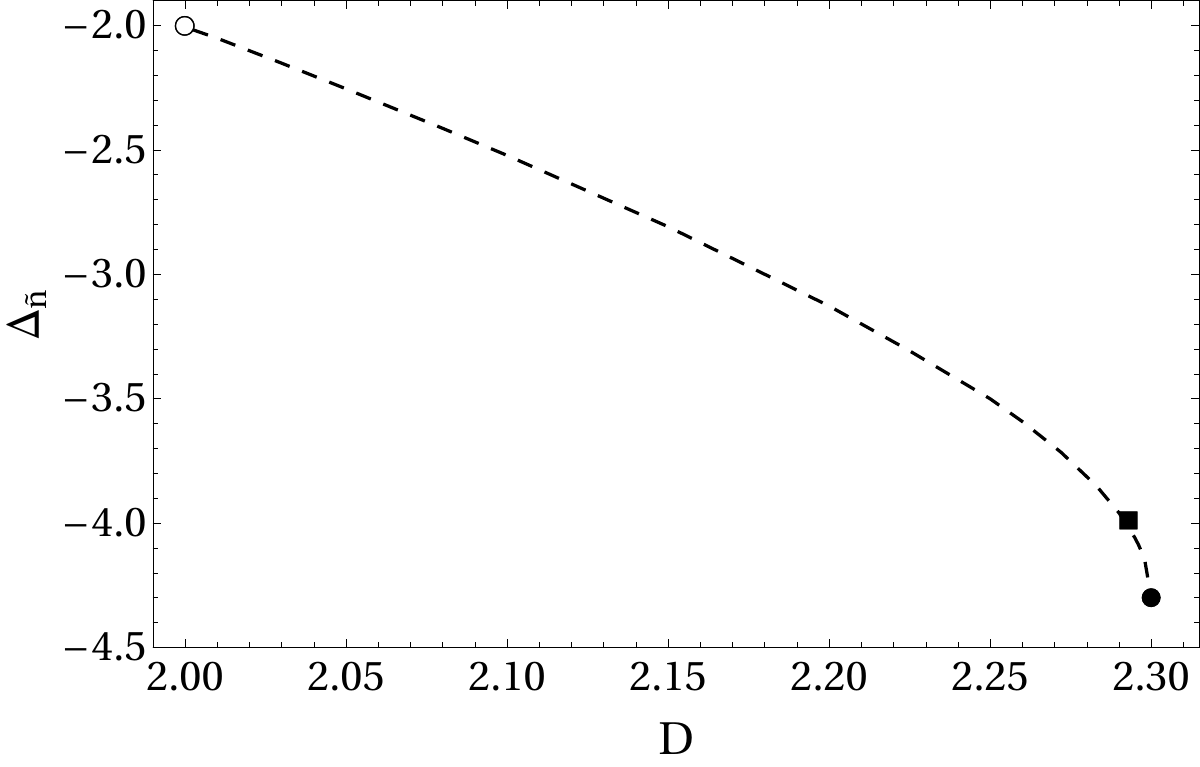}
\caption{ Scaling dimension coefficient of the momentum distribution for three resonant identical atoms embedded in the unatomic spatial region. The black point denotes the critical dimension ($D_c = 2.3$), where the transition between continuum and discrete scaling symmetry takes place. The $D=2$ limit is denoted by the empty black circle and the black square represents the dimension in which an exchange between the leading and sub-leading order terms in the asymptotic expansion occurs, precisely at $\overline{D}=2.292$. }
\label{fig5}
\end{figure}

Below the critical dimension, the large  momentum behaviour of the trimer wave function is given by equation~\eqref{eq:unatomictail}. The single particle momentum distribution in this regime behaves like $\sim 1/q^{D+2+2s_1}$, which is the case for the term normalized by the three-body contact~\eqref{c2c3}. The scaling coefficient for this quantity is given by  
\begin{equation}
n(\lambda\ q) \rightarrow \lambda^{\Delta_{{\tilde{n}}}} n( q) \,,
\end{equation}
that depends on the effective dimension in which the system is embedded $\Delta_{{\tilde{n}}} \equiv -(D+2+2s_1)\,$. 

Fig~\ref{fig5} shows the scaling coefficient for the leading term in the unatomic region. The black point denotes the critical dimension where the transition from discrete to continuum scale symmetry occurs. At the dimension $\overline{D} = 2.292$, indicated by a black square, where  $\Delta_{\tilde{n}}=-4$, there is an exchange between the leading order term in the asymptotic expansion of $n(q)$~\ref{eq:nasympdbardc}. For dimensions below $\overline{D}$, the behavior of the system belongs to the unatomic one, where only the continuous scaling term is present.  

We can compare the asymptotic momentum distribution considering the system embedded in different spatial configurations. In the Efimov region, $D_c <D\leq 3$, one can write~\cite{Nb_AAB_Ddim_Efimov}
\begin{eqnarray}
n(q) \underset{q  \gg \kappa_0}{=} \frac{C_{2}}{q^{4}}  +
\frac{C_{3}}{q^{D+2}} \cos\!\left[2 s_0 \log\left(q/\kappa_0
  \right) \! + \!\phi \right] + \cdots \,,
\label{c3c3l}
\end{eqnarray} 
where the non-oscillatory behavior is driven by $C_2$ and the term with the characteristic Efimov log-periodic oscillation has amplitude weighted by $C_3$. Comparing equation~\ref{c3c3l} with equations~\ref{eq:nasympdbardc} and \ref{c2c3}, we observe that while $C_2$ weights $1/q^{4}$ and scales as $\kappa_0^{4-D}$ for $\overline{D}<D\leq3$, the three-body parameter $C_3$ weights $1/q^{D+2+s_n}$ for $2<D\leq 3$
and scales as $\kappa_0^{2}$ in the Efimov regime, and as $\kappa_0^{2-2s_1}$ in the unatomic regime. Experimentally, this is an important feature in order to fit the asymptotic momentum distribution data to finally extract the contact parameters. 

\begin{figure*}[t!]
\includegraphics[width=8.5cm]{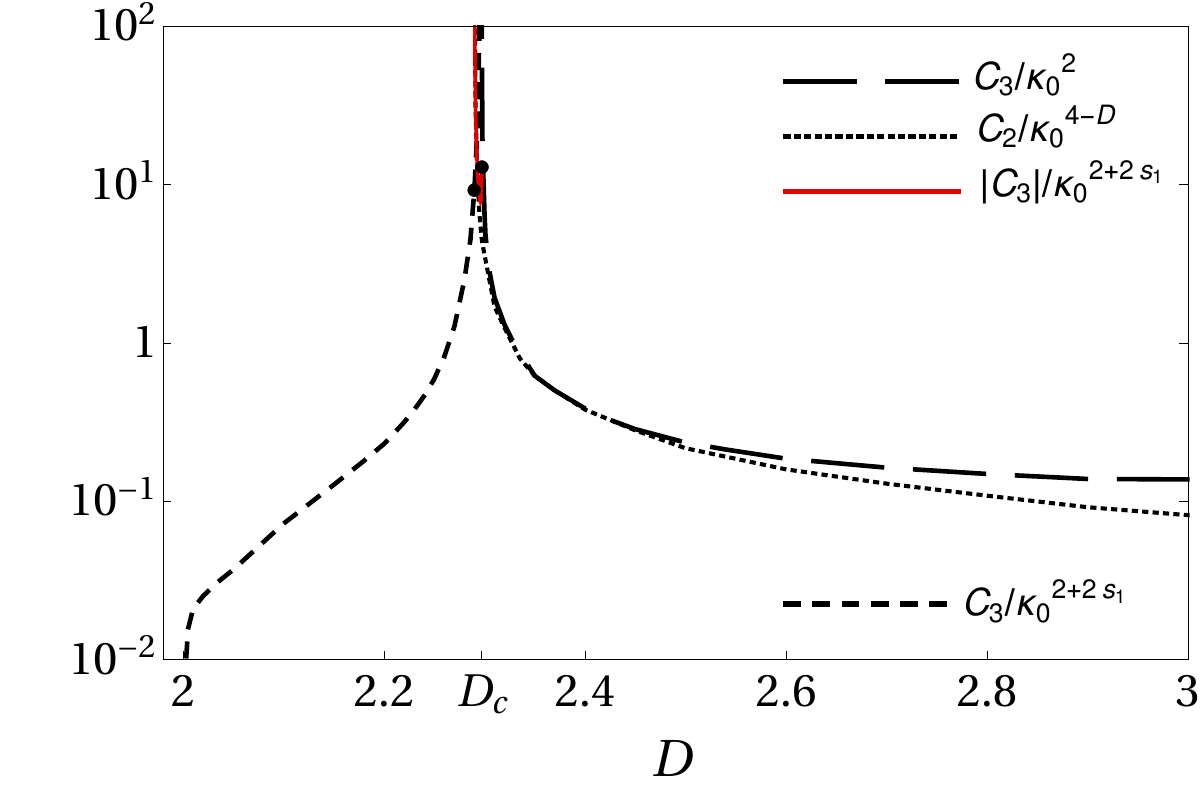}
\includegraphics[width=8.5cm]{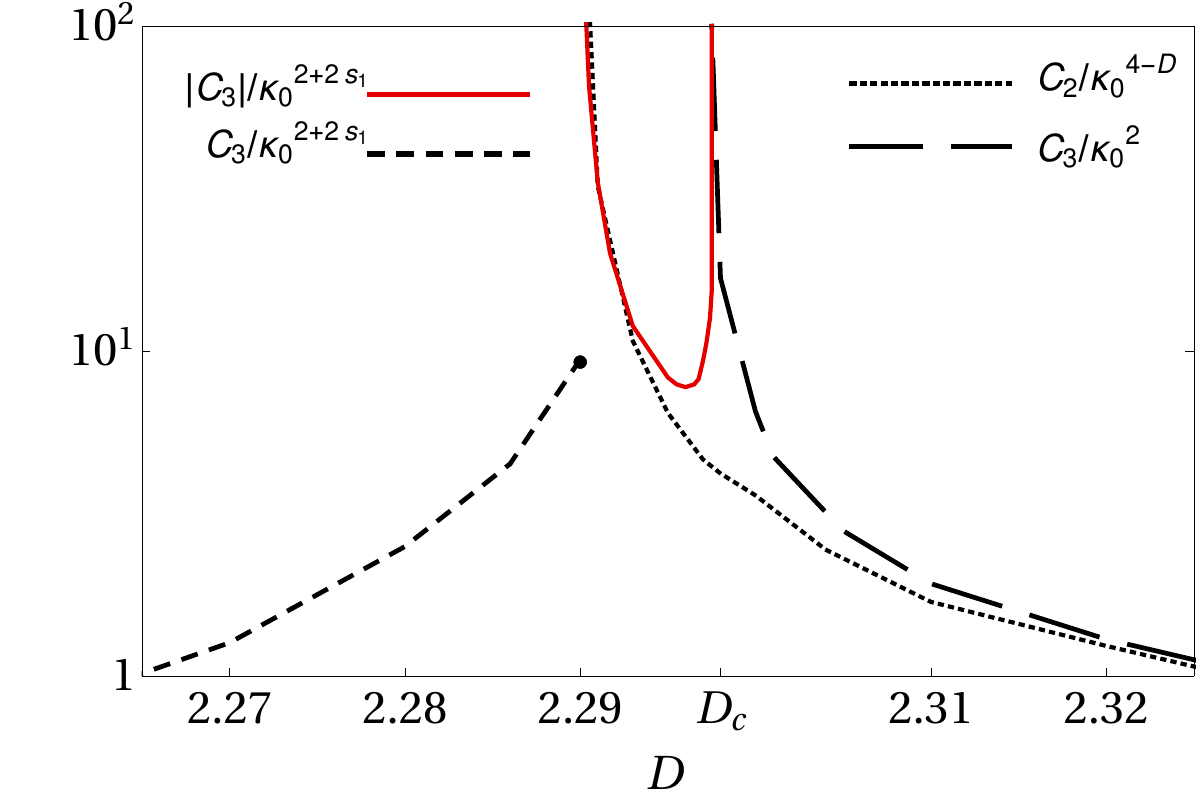}
\caption{Two- and three-body contact parameters  as a function of the noninteger dimension. The crossover from Efimov to the unatomic regime takes place at the critical dimension $D_c=2.3$. The right panel is a blow-up of the left panel around~$D_c$.}
\label{fig6}
\end{figure*}

In two-dimensions, the asymptotic momentum distribution has the form~\cite{Bellotti:2012dv}
\begin{equation}\label{eq:n2d}
    n(q)= \frac{C_{2}}{q^{4}}  +
\frac{\log^3(q)}{q^{6}}\,C_{3}+\cdots  .
\end{equation}
which can not be achieved with our analytical treatment, strictly applied to  $D>2$  as the system approaches the three-body threshold, finite two-body energy corrections has to be implemented. 

The key point of this work is the evolution of the trimer properties in the transition between the Efimov to the unatomic regime.
The findings that $C_2$ is not only present in the Efimov regime, but also in the unatomic region, for noninteger dimensions $\overline D<D<D_c$, precisely in the small interval $2.292<D<2.3$, presents a relevant feature for investigations ultracold atomic experiments. 

Fig.~\ref{fig6} shows, on the left side, the dimensionless two and three-body contact parameters as a function of the noninteger dimension $D$ where the system is effectively embedded. In the interval $\overline D<D<D_c$, the two- and three-body contacts are present, but they are barely visible in this figure. At the critical dimension, where $s_n\equiv 0$, $C_3/\kappa_0^{2}$ shows a remarkable peak, suggesting a dramatic evolution of the momentum density when squeezing the system in the transition from discrete to continuous scale symmetry regimes. Furthermore, in the unatomic regime $C_3/\kappa_0^{2+2s_1}$ tends to zero when approaching 2D, as the term $1/q^{2}$ is absent in the 2D asymptotic density~\eqref{eq:n2d}. On the right side, Fig.~\ref{fig6} exhibits a blow-up of the left panel around $D_c$, where it is shown that the two-body contact diverges for $D=\overline D_+$, presenting a quite rapid variation in such a small interval. The contacts in the small range given by $\overline D<D<D_c$ have opposite sign and cancel each other at $\overline{D}$.  

\section{Summary}
\label{secVI}

In this work, we explored the theoretical possibility of creating unatomic trimers by squeezing an atomic trap from three to two dimensions with the $s$-wave dimer at the continuum threshold. The continuous spatial change of the trap, which drives the trimer from the Efimov to the unatomic regime, is mimicked by a noninteger dimension. The trimer energy eigenstate was obtained by solving the Faddeev equations with contact interactions, combined with the Bethe-Peierls boundary condition, considering an infinite scattering length. To exclude non-normalizable solutions, we analyzed how the Skorniakov and Ter-Martirosian equation in noninteger dimensions for finite trimer binding energies behaves with different values of scale parameter.

The unparticle nature of the bound three-boson system, characterized by continuous scale invariance, is manifested in several properties of the trimer state when the system is located below the critical dimension $D_c$ and above $D=2$. In contrast to the log-periodicity found in the Efimov regime, the unatomic wave function  manifests a power-law form at short distances, or, correspondingly, at large momentum. 

The characteristic exponentially damped long distance behavior of the trimer wave function depends on its three-body binding energy $E_3 = - \hbar^2\kappa^2/m$. Typically, it is given by $\sim 1/\kappa_0$, which increases by decreasing the binding. We calculated the momentum distribution and showed that it has an asymptotic power-law dependence for momenta much larger than the characteristic binding energy. Besides that, we established the dependence on the noninteger dimension of the momentum distribution tail and contrasted the results with the findings in Ref.~\cite{Nb_AAB_Ddim_Efimov} for the Efimov regime. Finally, we showed that the contacts present a sharp transition at the crossover from the Efimov to the unatomic regime, at the critical dimension $D = D_c = 2.3$. The signature of the transition between the two regimes is a well-pronounced  peak of the three-body contact, while the two-body contact is still finite in a very limited dimensional range given by $\overline{D}<D\leq D_c$. The three-body contact vanishes for a squeeze towards two dimensions, as signalized by the absence of a term $1/q^{2}$ in the tail of the momentum density, providing another fingerprint of the unatomic behaviour, before the flat situation is established in the trap. 

In principle, the trimer momentum distribution can be explored through advanced experimental techniques. The available technology associated  with cold traps allows the direct measurement of the dimer and trimer properties in atomic gases~\cite{fletcher0,musolino} and eventually extract their evolution in~$D$ by continuously squeezing the trap.

\section*{Acknowledgments} This work was partially supported by: Funda\c{c}\~ao de Amparo \`a Pesquisa do 
Estado de S\~ao Paulo (FAPESP) [grant numbers 2019/07767-1 (T.F.), 
2023/02261-8 (D.S.R.), 2018/25225-9 (G.K.) and 2023/08600-9 (R. M. F.)], Conselho Nacional de Desenvolvimento 
Cient\'{i}fico e Tecnol\'{o}gico (CNPq) [grant numbers 306834/2022-7 (T.F.), 
302105/2022-0 (M.T.Y.) and 309262/2019-4 (G.K.)] and Coordenação de Aperfeiçoamento de Pessoal de Nível Superior (CAPES) Grant number
88887.928099/2023-00 (D.S.R).  This work is a part of the
project Instituto Nacional de  Ci\^{e}ncia e Tecnologia - F\'{\i}sica
Nuclear e Aplica\c{c}\~{o}es  Proc. No. 464898/2014-5.

\appendix
\section{Momentum distribution tail}
\label{app:a}

In this appendix, we derive the large momentum equations in the Unatomic region for the three identical bosons system studied in this work. For a detailed derivation of these equations in the Efimov region, please see Ref.~\cite{Nb_AAB_Ddim_Efimov}.

 \subsection{$n_1(q)$}
\label{appn1}
 
 Eq.~\eqref{n1} is written in spherical coordinates as
\begin{equation}
 n_{1}(q) =  \lvert \chi(q) \rvert^{2} \mathcal{S}_{D} \int^\infty_0 \hspace{-.3cm}d p \frac{p^{D-1}}{\left(\kappa_0^2 + p^{2}+q^{2}/2\mu\right)^{2}}, 
 \end{equation}
 where $\mathcal{S}_D = 2\pi^{D/2}/\Gamma(D/2)$. Changing variables $p/q = y$ and considering $q\gg \kappa_0$, we have that
 \begin{multline}
 n_{1}(q) =  \frac{\lvert \chi(q) \rvert^{2}}{q^{4-D}}\mathcal{S}_{D} \int^\infty_0 d y \frac{y^{D-1}}{\left( y^{2} + 1/2\mu\right)^{2}}  \\
=\frac{\lvert \chi(q) \rvert^{2}}{q^{4-D}}\left(2\mu\right)^{2-\frac D2}\pi\mathcal{S}_{D} 
\left(\frac{1}{2}-\frac D4\right) \csc\left( \frac{D\pi}{2} \right)\,.
 \end{multline}
 
The large momentum tail of $n_1(q)$ is obtained using the asymptotic form of the spectator function, Eq.~\eqref{asympespectunatomic}, which leads to:
\begin{equation}\label{eq:A3}
 n_{1}(q)\underset{q  \gg \kappa_0}{\propto} \frac{\left(2\mu\right)^{2-\frac D2}\pi}{q^{D+2+2s_1}}
 \mathcal{S}_{D} 
\left(\frac{1}{2}-\frac D4\right) \csc\left( \frac{D\pi}{2} \right)\,.
 \end{equation}

  \subsection{ $n_2(q)$}
\label{appn2}

Taking the large momentum limit, where $q\gg \kappa_0$, the second contribution, $n_2(q)$, can be computed 
from Eq.~\eqref{n2}. Making the change of variables $\textbf{p}
-\textbf{q}/2=\textbf{v}$, we find
\begin{equation}
 n_{2}(q) \underset{q  \gg \kappa_0}{=} 2 \int d^{D}v \frac{\lvert \chi(v) \rvert^{2}}
 { \left(  v^{2}+ \textbf{v}.\textbf{q} + q^{2}  \right)^{2} }\, .
 \end{equation}
In order to identify the leading order term in the large momentum region for $D>\overline D$, we perform the 
manipulation
\begin{eqnarray}\label{eq:n2}
n_{2}(q) &\underset{q  \gg \kappa_0}{=} &2 \int \hspace{-.1cm} d^{D}v\lvert \chi(v) \rvert^{2}\left[ \frac{1}
 { \left( v^{2}+ \textbf{v}.\textbf{q} + q^{2}   \right)^{2} }-\frac{1}{q^4}\right] \nonumber \\
 &+&\frac{C_2}{q^{4}},
 \end{eqnarray}
where $v\equiv|\textbf{v}|$ and $C_2$ is the two-body contact, given by
\begin{eqnarray}
\hspace{-.1cm} C_{2}
&=& 2 \int^\infty_0 \hspace{-.1cm}d^Dv \, \lvert \chi(v) \rvert^{2}. 
\label{eq:c2}
 \end{eqnarray}
Changing variables to $y=v/q$, using hyper spherical coordinates and integrating over the angle, we have that
\begin{equation}\label{eq:n2C2}
\hspace{-.2cm} n_{2}(q) \underset{q  \gg \kappa_0}{=}\frac{C_2}{q^{4}}+ \frac{2\mathcal{S}_D}{q^{4-D}} \int dy \ y^{D-1}\lvert \chi(q\ y) \rvert^{2}\left[ \mathcal{H}(y)-1\right] 
 \, ,
 \end{equation}
 where
\begin{multline}
 \mathcal{H}(y)\equiv\frac{D-2}{ y^4+y^2+1} 
 +\frac{(3-D) \left(  y^2+1\right) \, }{[  (y-1) y+1]^2 [  y (y+1)+1]} \\
 \times H_2F_1\left(1,\frac{D-1}{2},D-1,-\frac{2  y}{ (y-1) y +1}\right),
\label{eq:H2F1}
 \end{multline}
with $H_2 F_1(a,b,c,z)$ denoting the regularized hyper-geometrical function that for $y\to0$ is given by 
\begin{equation}\label{eq:expH}
 \mathcal{H}(y)\underset{y  \rightarrow 0}{=}  1+\frac{3-2D}{D}y^2+\mathcal{O}[y^3]\,.
\end{equation}

Using the asymptotic form of the spectator function in the unatomic region, 
the subleading contribution to the density can be written as
 \begin{eqnarray} n_{2}(q)-\frac{C_2}{q^4} &\underset{q  \gg \kappa_0}{\propto} & { 2 \,\mathcal{S}_D\over q^{D+2+2s_1}} \int^\infty_0 \hspace{-0.1cm}dy \,  y^{1-D-2s_1}\nonumber  \\
&\times&\left[ \mathcal{H}(y)-1\right], \label{eq:B4}
 \end{eqnarray}
 this expression presents an ultraviolet divergence at
 $D=\overline D=2-2s_1$ ($\overline D=2.292$ for  identical bosons), and as long
\begin{equation}
D>\overline D  \, , 
\end{equation}
which corresponds to $\Delta_{n}=-D-2s_1-2>-4 $, it is finite. The UV divergence at $D=\overline D$ is  due to the term having the -1 in the integral of Eq.~\eqref{eq:B4}, while the term with $\mathcal{H}(y)$ is UV finite and due to its expansion around $y=0$,  Eq.~\eqref{eq:expH}, the infrared divergence from the term -1 appearing when approaching $D\to\overline D_+$ is canceled. 

One should note that for $D\to\overline D_+$ ($\overline D=2-2s_1$) the density $n_2(q)$ is finite, although $C_2$  and the contribution to $C_3$ diverges. The reason for the cancellation  can be easily seen by noticing that  $C_2\to -C_3\to \infty$, and thus the divergent contribution of the two-body contact is exactly canceled by the three-body contact, letting the tail of $n_2$ finite.

For $2<D<\overline D$, the formula for $C_2$, Eq.~\eqref{eq:c2}, is not valid because the factorization used to define the two-body contact cannot be done due to the UV divergence, and consistently only remains in $n_2(q)$ the contribution to the three-body contact, which now is the leading one
\begin{eqnarray}\label{eq:n2C3}
 n_{2}(q) \propto \frac{2\mathcal{S}_D}{q^{D+2+2s_1}} \int^\infty_0 dy \  y^{1-D-2s_1}\, \mathcal{H}(y) \, .
 \end{eqnarray}
We should observe that when approaching $D\to \overline D$, the above expression has an IR divergence, and instead the calculation of the asymptotic form of $n_2(q)$ has to be performed with
\begin{equation}\label{eq:n2C3a}
\hspace{-.2cm} n_{2}(q) = \frac{2\mathcal{S}_D}{q^{4-D}} \int dy \ y^{D-1}\lvert \chi(q\ y) \rvert^{2} \mathcal{H}(y) 
 \, ,
 \end{equation}
using the spectator function in equation~\ref{regulatespect} which is free of UV and IR divergences. In practice, the asymptotic value of $q^{D+2+2S_1}n_2(q)$ for $D$  approaching $\overline D_-$ appears at larger and larger values of the momentum, and exactly in the limit of $D\to\overline D_-$, one has that $q^{4}n_2(q)$ is finite for $q\to \infty$.

 \subsection{ $n_3(q)$}
\label{appn3}
 
Taking $n_3(q$) from Eq.~\eqref{n3} with the change of variables $\textbf{v}=\textbf{p} -\textbf{q}/2$ and considering the large momentum limit, namely $q\gg \kappa_0$, we can write that
  \begin{equation}
n_{3}(q)\underset{q  \gg \kappa_0}{=}  \int d^{D}v \frac{2\chi^*{(}q)\ \chi(v )  }
 { \left( v^{2} +\textbf{v}\cdot\textbf{q} +q^{2}  \right)^{2} } + {\rm c.c.}\,.
 \end{equation}
The spectator function is real, changing variables $v/q =y$ and integrating over the angles in spherical coordinates, we get that
 \begin{equation}
 n_{3}(q)\underset{q  \gg \kappa_0}{=} \chi(q)\frac{4\mathcal{S}_D}{q^{4-D}} \int^\infty_0 \hspace{-.3cm}  dy  \ y^{  D-1}  \chi(q\, y)\,\mathcal{H}(y)\,,
 \end{equation}
where $\mathcal{H}(y)$ is  given by Eq.~\eqref{eq:H2F1}. The asymptotic form is found by using  the spectator function  tail from Eq.~\eqref{asympespectunatomic}, leading to
\begin{eqnarray}
 n_{3}(q)\underset{q  \gg \kappa_0}{\propto}\frac{ 4\mathcal{S}_{D}}{q^{D+2+2s_1}} \int^\infty_0 dy \ y^{ -s_1}  \ \mathcal{H}(y)\,. 
 \end{eqnarray}

\subsection{ $n_4(q)$}
\label{appn4}

The  arguments of the spectator functions in  Eq.~\eqref{n4} are
  \begin{equation}
 \lvert \textbf{p}\pm\frac{\textbf{q}}{2}\rvert = q \, y_{\pm}, \quad \text{where}\quad y_{\pm}=\sqrt{y^{ 2}+\frac14\pm y\cos\theta}\,,
 \end{equation}
with  $y=p /q$. Considering the large momentum limit, one has
\begin{eqnarray}\label{eq:D2}
 n_{4}(q) &\underset{q  \gg \kappa_0}{=}& \frac{4\pi}{q^{4-D}} \prod_{k=1}^{D-3}\int_0^{\pi}d\theta_k  (\sin\theta_k)^{k}\int^\infty_0 dy  \frac{y^{ D-1}  }
 { \left( y^{ 2} + \frac34\right)^{2} }\nonumber   \\
 &\times& \int_0^{\pi}d\theta \,(\sin\theta)^{D-2} \,\chi^*{(} q \, y_{ -})\chi(q \, y_{ +})\, ,\ \ \ \ \
 \end{eqnarray}
The product of the two spectator functions  tails from Eq.~\eqref{asympespectunatomic} allow us to write that: 
\begin{small}
\begin{eqnarray}\label{eq:D5}
 n_{4}(q)\underset{q  \gg \kappa_0}{\propto}\frac{1}{q^{D+2+2s_n}} \frac{4\pi^{\frac{D-1}{2}}}{\Gamma\left(\frac{D-1}{2}\right)} \int^\infty_0 dy \ \frac{y^{ D-1}  }
 { \left( y^{ 2} + \frac34 \right)^{2} }\ \ \ \ \ \ \ \ \ \ \ \ 
 \nonumber \\
 \times \int_0^{\pi}d\theta (\sin\theta)^{D-2}\Big[\Big(y^2+\frac14\Big)^2-y^2(\cos\theta)^2\Big]^\frac{1-D-s_1}{2} \,. \ \ \ \
 \end{eqnarray}  
\end{small}

\end{document}